\documentclass[sigconf,edbt]{acmart-edbt2020}

\def\BibTeX{{\rm B\kern-.05em{\sc i\kern-.025em b}\kern-.08em
    T\kern-.1667em\lower.7ex\hbox{E}\kern-.125emX}}

\usepackage{url}
\usepackage{enumerate}
\usepackage{graphics}
\usepackage{graphicx}
\usepackage{multicol}
\usepackage{multirow}
\usepackage{float}
\usepackage{subfigure}
\usepackage{comment}
\usepackage{graphicx,wrapfig}
\usepackage{amsfonts}
\usepackage{todonotes}
\usepackage{mathtools}
\usepackage{graphicx}
\usepackage{caption}
\usepackage[title]{appendix}
\usepackage{listings}
\usepackage{amsmath}
\usepackage{color}
\usepackage{textcomp}
\usepackage[linesnumbered,vlined, ruled]{algorithm2e}

\usepackage{booktabs} 

\setcopyright{rightsretained}

\begin{document}
\title{OBDA for the WEb: Creating Virtual RDF Graphs \\ On Top of Web Data Sources \large{[Application Papers]}}

\author{Konstantina Bereta, George Papadakis, Manolis Koubarakis}
\affiliation{%
  \institution{National and Kapodistrian University of Athens, Greece}
}
\email{{Konstantina.Bereta, gpapadis, koubarak}@di.uoa.gr}

\renewcommand{\shortauthors}{K. Bereta et al.}

\begin{abstract}
Due to Variety, Web data come in many different structures and formats, with HTML tables and  REST APIs (e.g., social media APIs) being among the most popular ones. A big subset of Web data is also characterised by Velocity, as data gets frequently updated so that consumers can obtain the most up-to-date version of the respective datasets.
At the moment, though, these data sources are not effectively supported by Semantic Web tools. To address variety and velocity, we propose \textsf{Ontop4theWeb}, a system that maps Web data of various formats into virtual RDF triples, thus allowing for querying them on-the-fly without materializing them as RDF. We demonstrate how \textsf{Ontop4theWeb} can use SPARQL to uniformly query popular, but heterogeneous Web data sources,
like HTML tables and Web APIs. We showcase our
approach in a number of use cases, such as 
 Twitter, Foursquare, Yelp and HTML tables. 
We carried out a thorough experimental evaluation which verifies the high efficiency of our framework, which goes beyond the current state-of-the-art in this area, in terms of both functionality and performance.
\end{abstract}

\maketitle

\section{Introduction}
\label{sec:intro}

Querying Web data sources on-the-fly is an important task for several reasons: 
(i) Having full access to such data sources may involve a high economic cost (e.g., the price of subscribing to the entire Twitter stream). 
(ii)The constantly changing terms of use and the corresponding legislation complicates data crawling (e.g., the constraints defined by the recent EU General Data Protection Regulation\footnote{See \url{https://eugdpr.org} for more details.}).
(iii) The high frequency of updates (\textbf{Velocity}) makes it difficult for data consumers to keep up with the content published in popular Web sources like social media applications. For example, in Twitter, 
approximately 6.000 tweets are posted per second\footnote{\url{http://www.internetlivestats.com/twitter-statistics/}}.

Moreover, \textit{querying on-the-fly non-RDF Web data using SPARQL} has become a major issue \cite{ISWC18,ldow18,eswc18}, because 
many Web data sources rely on non-RDF formats, such as REST APIs and HTML tables. To address this \textbf{Variety},  SPARQL is extended in \cite{eswc18} so that it allows for querying RDF data in combination with data coming from Web APIs in the form of JSON files. In \cite{ldow18}, the authors propose an architecture based on micro-services that extends the SPARQL protocol with the ability to query APIs on-the-fly. Finally, an extension of the R2RML mapping language is proposed in \cite{ISWC18}, providing primitives for querying various kinds of Web data sources, such as APIs. 

However, these works merely support relational data or specific file formats (e.g., XML, CSV). They also rely on custom SPARQL/R2RML extensions that hamper their adoption, while combining them with third-party added-value services is a very complicated procedure. Some of them also implement a caching mechanism \cite{eswc18,ldow18}, but are inherently incapable of making the most of it, 
as demonstrated by our experiments in Section
\ref{sec:evaluation}.

\begin{figure}[t]
   \begin{center}
   		\includegraphics[scale=0.15]{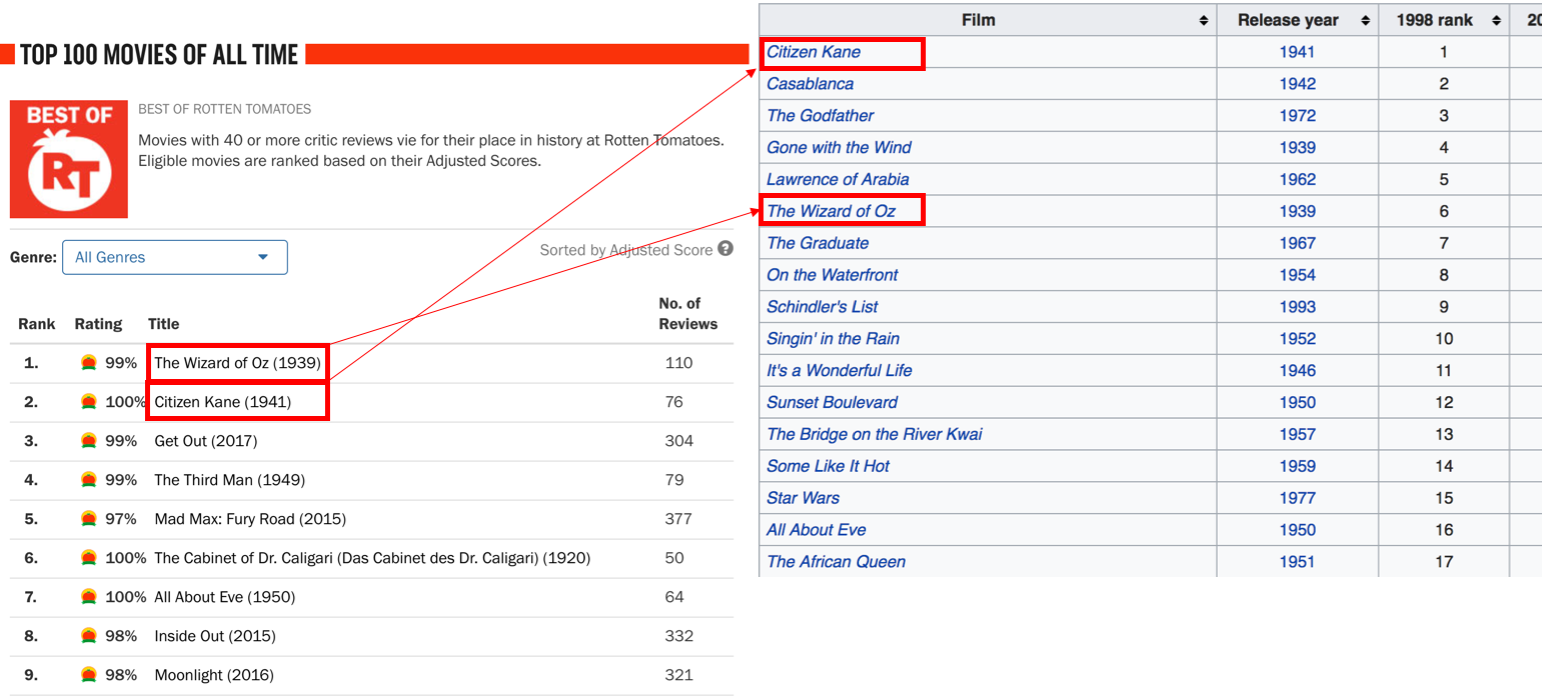}
   		\vspace{-4pt}
   	    \caption{Tables with 100 movies from Rotten Tomatoes and Wikipedia.}
   	    \vspace{-5pt}
        \label{fig:films}
   \end{center}
\end{figure}

For example, let us assume that we would like to query data available in the HTML
tables shown in Figure \ref{fig:films} using SPARQL. The first table presents
a listing of movies with the best rank, provided by RottenTomatoes\footnote{\url{https://www.rottentomatoes.com/}}. The second one is a Wikipedia table that lists the best 100 movies. If we would like to query this data using SPARQL, 
we would have to:
\begin{enumerate}
    \item create a custom parser to parse the data
    \item convert the data into RDF
    \item store the RDF data in a triple store, and eventually, 
    \item query the data using SPARQL.
\end{enumerate}
The only alternative approach would be to
convert and store the data into relational tables, instead of using a triple store, and then
use an OBDA or RDB2RDF system to query the data using SPARQL. In both cases, 
the convert-and-store tasks could be inevitable. The approach presented in \cite{eswc18} 
could not be applied, as it is not desinged for HTML tables.

In this paper, we go beyond these state-of-the-art works,
introducing a system, called \textsf{Ontop4theWeb}, that extends existing ontology-based data access (OBDA) techniques to support uniform queries over data from different Web sources, such as WebTables and REST APIs. \textsf{Ontop4theWeb} relies on virtual relational tables that allow for executing any SPARQL query on top of an OBDA system, using the necessary ontology and mappings, but without requiring the data to be available a-priori, i.e., before the query is posed. In this way, \textsf{Ontop4theWeb} offers a series of unique characteristics: 
\begin{enumerate}
    \item It accommodates any format of Web data, such as the increasingly popular HTML tables and the omnipresent REST APIs, and it is especially suitable for data sources with high Velocity.
    \item It is easy to use and incorporate into any Semantic Web application, as it relies on standard SPARQL and R2RML (and its equivalents).
    \item Based on micro-services, it facilitates the seamless enrichment of retrieved data with third-party added-value services (e.g., sentiment analysis).
    \item It is able to fully exploit an effective caching mechanism that can be configured in line with the update rate of the respective data source.
    \item It goes beyond traditional convert-and-store approaches by requiring no materialisation of the original data following the new schema. 
\end{enumerate}

Overall, the main contributions of this paper are as follows:
\begin{itemize}

\item We introduce \textsf{Ontop4theWeb}, an OBDA-based system for posing SPARQL queries on top of non-RDF Web data on-the-fly, i.e., they are fetched at query time, rather than importing or downloading them a-priori. To achieve this, virtual table operators are embedded in the SQL queries that are included in R2RML mappings. These mappings specify which part and source of Web data will be fetched and how they will be mapped to virtual RDF terms. Combining these mappings with an ontology allows for returning the virtual relational data that are involved in the query as RDF results.

\item We showcase the applicability of \textsf{Ontop4theWeb} in three use cases that: (i) involve significant amount of heterogeneous crowd-sourced information (Variety), (ii) get updated so frequently that a snapshot of the respective information at a given time might become outdated soon (Velocity), and (iii) are widely used by application developers.

\item We experimentally evaluate \textsf{Ontop4theWeb}, demonstrating its feasibility and scalability in the three, realistic, highly diverse and demanding applications we consider. The results show that our approach is able to process queries on WebTables of up to 100,000 rows in size within minutes. We also compared the performance of our approach to the state-of-the-art method described in \cite{eswc18}, with the results verifying that our framework provides more functionality, while being more efficient, as well.
\end{itemize} 

The rest of the paper is organized as follows:
in Section \ref{sec:related}, we discuss the state-of-the-art in the field, 
while in Section \ref{sec:preliminaries}, we briefly present basic background knowledge.
Section \ref{sec:approach} describes our approach and methodology, whereas
Section \ref{sec:implementation} documents our system, which is applied to practical use cases in Section \ref{sec:usecases}. Section \ref{sec:evaluation} presents
our experimental evaluation, with Section \ref{sec:conclusions} concluding the paper along with directions for future work.
\section{Related Work}
\label{sec:related}

OBDA systems \cite{poggi} are primarily useful in cases where users store their data in  relational databases, but do not want to materialize them as RDF triples, particularly when these databases are large or/and get frequently updated (Velocity) \cite{mastro}. As a result, many OBDA and RDB2RDF systems have been developed in the recent years, such as Ontop \cite{ontop}, Ultrawrap \cite{ultrawrap}, Morph \cite{morph}, Sparqlify\footnote{\url{http://aksw.org/Projects/Sparqlify.html}}, and Oracle Spatial and Graph\footnote{\url{http://www.oracle.com/technetwork/database/options/spatialandgraph/overview/index.html}}. These systems are able to connect to \emph{existing} relational data sources and create virtual RDF graphs using ontologies and mappings. The common assumption of these systems is that the source data are materialized and connection details are provided in the mappings. Most of them support the R2RML mapping language or provide translators from their native mappings languages to R2RML. For example, Ontop also supports its own native OBDA language for encoding mappings. Once connected to the data source, OBDA systems make the most of the underlying database by collecting information about data characteristics (e.g., statistics, constraints).

On another line of research, there are RDB2RDF systems that focus on converting data into RDF using mappings to produce RDF dumps. Initially, only relational data sources were supported through the R2RML language \cite{rtorml}. Given, though, that data can be found in many formats other than relational, the RML language was created as a superset of R2RML, encoding how various data formats, like XML and CSV, can be mapped to RDF triples \cite{rml}. Another recent work in this direction is the approach described in \cite{sparql-generate}, which aims at converting Web data from various formats (e.g., CSV, JSON) into RDF,
using SPARQL~queries - SPARQL 1.1 primitives and extension functions were extended, too.

Recently, a mapping language, called D2RML, was proposed in \cite{ISWC18}, inspired from R2RML and RML. This work extends R2RML to support more data formats, including REST APIs. Although an implementation of a D2RML processor  exists, it is not part of a standalone SPARQL query engine, to the best of our knowledge.

Closer to our work is the \textit{SERVICE-to-API} system \cite{eswc18}, which proposes an extension of SPARQL that enables users to  combine the responses of JSON APIs with results from the evaluation of standard triple patterns. We deviate from this approach in that:
\begin{enumerate}
    \item we do not extend SPARQL syntax, 
    \item  we allow users to query APIs using standard SPARQL triple patterns directly,
without having to combine them with stored RDF data,
    \item we provide a general approach that is not limited to JSON APIs,
    \item  we produce significantly fewer API calls, which translates to improved performance. 
\end{enumerate}
Section \ref{sec:evaluation} provides a detailed qualitative and quantitative comparison between the two systems.

Also close to our work is the architecture proposed in \cite{ldow18}, which is based on the development of SPARQL wrappers for Web APIs. To this end, it extends HTTP requests to SPARQL endpoints to include arguments that are used to retrieve a fragment of the data that can be accessed via the Web API. This fragment is converted into RDF and stored using an in-memory triple store. In this way, the SPARQL query that is contained in the original SPARQL HTTP request is evaluated against the RDF graph that is stored in the triple store, which is only a fragment of the original dataset. This fragment can be considered as a linked data fragment (LDF) interface, as described in \cite{ldf}. Note that the original linked data fragment approach considers the evaluation of single triple patterns on the server-side, leaving the rest to the client, in order to improve the sustainability of linked data endpoints \cite{ldf}. However, there is no limit to the expressivity of queries that can be executed on the server in \cite{ldow18}. In short, \cite{ldow18} converts a fragment of the dataset into RDF and stores the converted data into an in-memory triple store. In our approach, the conversion is performed on-the-fly using mappings and an in-memory virtual table is constructed instead.

We deviate from this approach in that: 
\begin{enumerate}
    \item no data are materialized into RDF, as the query is converted on-the-fly using mappings,
    \item our approach can be adapted to a different schema simply by modifying the mapping file, without requiring a change in the system code, as in \cite{ldow18},
    \item the translation of the original SPARQL query is completely transparent to the end-user, whereas \cite{ldow18} requires the end user to be fully aware of the Web API documentation so as to specify the fragment of the Web API that needs to be accessed. 
\end{enumerate}

In the area of databases and data integration, \cite{data-integration} gives
an overview of how web sources can be accessed using wrappers. The approach described in this paper follows the same principles, as the architecture that we propose contains
two layers that can be considered as wrappers on top of Web data sources, as we describe later on. However, we target specifically the problem of how one can pose SPARQL queries on top of Web data sources using ontologies and mappings, and this problem goes beyond the approaches discussed in \cite{data-integration} (although ontologies are mentioned).
\section{Preliminaries}
\label{sec:preliminaries}

We now present the background knowledge that forms the basis for defining the problem we are tackling as well as \textsf{Ontop4theWeb}.

\subsection{RDF and SPARQL}
We denote as  $I$, $B$ and $L$ the pairwise disjoint infinite sets of IRIs, blank nodes and literals, respectively. $V$ stands for the infinite set of
variables that are disjoint from $I$, $B$ and $L$. Based on \cite{SPARQL}, we provide the following definitions:

\noindent \textit{RDF triple}. An RDF triple is an element of the form $(s,p,o)$  of $(I\cup B) \times I \times (I \cup B \cup L)$, where
$s$ is the subject, $p$ is the predicate and $o$ is the object.

\noindent \textit{RDF graph}. An RDF graph is  finite set of RDF triples. 

\noindent \textit{Triple pattern}. A triple pattern is an element
of the form $(I \cup B \cup V) \times (I \cup V) \times (I \cup B \cup L \cup V).$

\noindent \textit{Graph pattern}. A (basic) graph pattern (BGP) is a finite set of triple patterns. 


\noindent \textit{Evaluation of triple patterns over an RDF graph}. 
Let $D$ be an RDF graph over $I \cup B \cup L$, t a triple pattern and $P_1,P_2$ graph patterns. The mapping $\mu$ is a partial function $\mu : V \mapsto (I \cup B \cup L)$, while $\mu(t)$ is the triple obtained if we replace every variable $u$
of the variables included in $t$ (i.e., $var(t)$) with their bindings according to $\mu$ (i.e., $\mu(u)$). In this context, the evaluation of a graph pattern over $D$, denoted by $[[.]]_D$, is recursively defined as follows: 
\newcommand{\leftsemijoin}{\mbox{$\mathrel{\raise1pt\hbox{\vrule height2.5pt depth0pt width0.6pt\hskip-1.5pt$>$\hskip -2.5pt$<$}}$}}

\begin{itemize}
  \item $[[t]]_D = \{\mu | dom(\mu) = var(t) \text{ and } \mu(t) \in D\}$, where $var(t)$ is a set of variables occurring in $t$.
  \item $[[(P_1 \text{ AND } P_2)]]_D = [[P_1]]_D \Join [[P_2]]_D$.
  \item $[[(P_1 \text{ OPT } P_2)]]_D = [[P_1]]_D \text{ } \leftsemijoin \text{ } [[P_2]]_D$.
  \item $[[(P_1 \text{ UNION } P_2)]]_D = [[P_1]]_D \cup [[P_2]]_D$.
\end{itemize}

\subsection{Ontology-based data access (OBDA)}
The OBDA paradigm \cite{obda} 
proposes the creation of virtual RDF graphs on top of
relational databases using ontologies and mappings. 
Given a database schema $S$, an ontology $O$, and a set of mappings $M$, an OBDA specification is defined as $P=(O,M,S)$. 
Then, an \emph{OBDA instance} $(P,D)$ is defined given
the OBDA specification $P$ and the database $D$ that follows
the database schema $S$. \textit{Mappings} encode how relational data get mapped into RDF terms. A virtual RDF graph $VG_{M,D}$ of the database instance $D$ is
produced if we apply the mappings $M$ to $D$. Then, if $[[Q]]_{(P,D)}$ is the evaluation of the SPARQL query $Q$ over the OBDA instance $(P,D)$, it is equivalent to $[[Q]]_{(P,VG_{(M,D)})}$.

The W3C standard language for encoding mappings is R2RML \cite{rtorml}.
As an example, we can map a relational table named \texttt{Student} with columns \texttt{id} and \texttt{name} into RDF with the following R2RML mapping:

{\footnotesize
\begin{verbatim}
<r2rml_mapping>  a rr:TriplesMap;
rr:logicalTable [ rr:sqlQuery """select id, name from Students""" ];
rr:subjectMap [ rr:template ex:{id}; rr:class ex:Student ];
rr:predicateObjectMap [ rr:predicate 	ex:hasName;
rr:objectMap [ rr:column "name" ; rr:datatype xsd:String ]].
\end{verbatim}
}

Yet, many OBDA systems support their own native mapping languages. In this work, we use the native language of Ontop \cite{ontop}, as it combines brevity with readability. The equivalent to the above R2RML mapping is the following:

{\footnotesize
\begin{verbatim}
[MappingDeclaration] 
[[mappingId obda_mapping
  target	ex:{id} a ex:Student ; {name}^^xsd:String .
  source	select id, name from Students ]]
\end{verbatim}
}

In this mapping, the target part provides templates of virtual triples that are generated using the values of these columns, while
the source part contains an SQL query that retrieves
the id and name columns of the table Students.




\section{Approach}
\label{sec:approach}

We now elaborate on how we extend SQL with virtual table operators that access heterogeneous Web data sources (Section \ref{subsec:sql}), on how we evaluate standard SPARQL queries on top of Web APIs (Section \ref{subsec:semantics}) as well as on the steps comprising our methodology (Section \ref{subsec:methodology}).

\subsection{Extending SQL with virtual table operators}
\label{subsec:sql}

The core concept of our approach is to model a data source
as a virtual relational table. For this reason, we define a virtual
table operator for each kind of data source. Each \textit{virtual table
operator} has the syntax:
$\mathbf{VT ::=  vtable (args[\ldots ,f])}$,
where the vector $args$ denotes the arguments that are given as input to the virtual table
operator, while 
$f$ is optional, denoting the \textit{cache update rate}. 

To understand the form of the SQL queries that use virtual tables, consider the extension of the SQL syntax in Listing \ref{lst:sqlsyntax}. 
\vspace{-2pt}
\begin{lstlisting}[language=SQL,basicstyle=\ttfamily\scriptsize,frame=tb,caption={SQL syntax for virtual tables},label={lst:sqlsyntax}]
<query specification> ::= SELECT [ <set quantifier> ] <select list>
                                    <table expression>
<table expression>    ::=  <from clause> 
                           [ <where clause> ] 
                           [ <group by clause> ]
                           [ <having clause> ]
<from clause>         ::= FROM <table references> 
<table references>    ::= <table reference> 
                           [ { <comma> <table reference> }..]  |
                           vtable_operator_name(args[,f])
\end{lstlisting}
\vspace{5pt}
We extend the SQL syntax provided in Listing \ref{lst:sqlsyntax} with virtual table support as shown in the last lines. The SQL 
standard defines two types of tables: (i) the \textit{base} ones, which are materialized in a database, and (ii) the \textit{derived} ones, which are
produced from relational algebra expressions.  
At the relational algebra level, a virtual table ($vtable$) is just another relational algebra operator. Thus, we consider virtual tables generated by virtual table
operators as another kind of derived~tables; any mapping language that 
is able to use SQL queries in mappings (e.g., R2RML, OBDA) is compatible.

To improve performance, each virtual table can optionally use a cache. The cache feature is useful in cases where:
\begin{enumerate}
    \item not all data sources get updated with the same frequency, 
    \item some data sources might not be accessible at the next query time (e.g., due to API limitations), 
    \item a minimal query execution time is required, due to a large number of queries, i.e., the frequency of queries is much higher than the update frequency of data sources.
\end{enumerate}

To support these cases, $f$ indicates the length of the time window
(in milliseconds),
during which the retrieved data are temporarily stored.
If the virtual table operator with the \textit{same} input parameters ($args$) is invoked
twice (or more) before this time window ends, the cached data will be used, improving query time. If 
the query is repeated after the end of the time window, the fresh data is fetched from the data source and gets stored in the system. If $f$ has a negative value or is completely absent, nothing is stored and the virtual table operator fetches fresh data every time it is invoked. To support this functionality,
we store meta-data about when and where data resulting from a virtual table signature was stored last time.

\begin{algorithm}[t]
{\small
\SetKwData{T}{T}\SetKwData{NOW}{NOW}\SetKwData{w}{w}
\SetKwData{t}{t}\SetKwData{row}{row}\SetKwData{e}{e}
\SetKwData{w'}{w'}\SetKwData{row}{row}\SetKwData{E}{E}
\SetKwFunction{getLastUpdate}{getLastUpdate}\SetKwFunction{getTableFromCache}{getTableFromCache}
\SetKwFunction{retrieveData}{retrieveData}\SetKwFunction{process}{processAttribute}
\SetKwFunction{getAttributes}{getAttributes}\SetKwFunction{UpdateCache}{UpdateCache}
\SetKwInOut{Input}{Input}\SetKwInOut{Output}{Output}
\Input{$args[\ldots ,f]$}
\Output{\T, the generated virtual table}
\Begin{
\T $\longleftarrow$ $\emptyset$\;
\t $\longleftarrow$ \getLastUpdate{$args$}\;
\If{$|t-$\NOW$| < f $}{   
 \T $\longleftarrow$ \getTableFromCache{$args$}\;
 \Return \T\;
 }
 $E$ $\longleftarrow$ \retrieveData{$args$}\;
\For{$e\in E$}{
    \row $\longleftarrow$ $\{tupleID\}$\; 
    $W \longleftarrow \getAttributes(e)$\;
        \For{$w \in W$}{
            \w'[] $\longleftarrow$ \process{\w}\;
            \row $\longleftarrow$ \row $\cup$ \{\w'[]\}\;
        }
    \T $\longleftarrow$ \T $\cup$ \{\row\}\;
    }
    \UpdateCache{$args$ \NOW, T}\;
    \Return{\T}\;
    }
}
\caption{Virtual Table Operator\label{vtable}}
\end{algorithm}

Since our approach and our caching mechanism deviates considerably from related works \cite{ldow18,eswc18}, we now explain in more detail how virtual tables work. Each virtual table operator is implemented differently, but a generalized description is provided in Algorithm \ref{vtable}. 
First, the operator checks the time the last query with the same arguments was executed (Line 3). If it is within the given cache update rate, $f$, the already retrieved results are returned as output (Lines 4-6). Otherwise, the operator retrieves the data from scratch, using the given arguments (Line 7). For each record, it creates a new tuple with a unique id (Lines 8-9). Next, it iterates over its attribute values, adding them to the tuple after the necessary processing (Lines 11-14). Note that the functionality of the \texttt{processAttribute} function ranges from simple tasks (e.g., data manipulation functions like value transformation/correction) to more complicated tasks (e.g., data mining tasks). For this reason, $w'$ may contain more than one element. E.g., it can be  the text of a tweet together with information about its polarity (i.e., positive or negative sentiment). Finally, after all tuples have been processed and added to the virtual table (Line 13), the cache is updated (Line 15) and the table is returned as output.

The result of a virtual table operator is a virtual table
with the following schema:
$\mathbf{VT [tupleID,{cols}]}$, where $tupleID$ is the unique identifier
of a tuple and $cols$ are the requested attributes. 
Note that some of these attributes might not exist originally
in the data source, but they could introduce new knowledge derived from
processing the original data, as shown in Section~\ref{subsec:twitter}.

\subsection{Evaluation of SPARQL queries on top of Web APIs}
\label{subsec:semantics}
As described above, the semantics of RDF and SPARQL \cite{SPARQL}
assume that the evaluation of SPARQL queries is performed over an
RDF knowledge base and the OBDA paradigm \cite{obda}  defines
the creation of virtual RDF graphs on top of materialised databases,
for which the schema is known a-priori. 

Our aim is to support the evaluation of SPARQL queries on top of 
different kinds of Web data (e.g., APIs, WebTables, etc.) without
extending SPARQL or the mapping
languages, as suggested by the related work \cite{ISWC18,ldow18,eswc18}. Instead, we extend the OBDA paradigm to support virtual relational data, for which the schema is not known a-priori, i.e., not before a SPARQL query is fired. 

Let us model the response of an API call as a set of 
sets of $<attribute,value>$ pairs. Let $ATTR$ be the set of all attributes of a response of an API call. For each $attr_{i} \in ATTR_{I} \subset ATTR$, we define a mapping $ \mu: attr_{i} \mapsto pred_{i}$ that maps $attr_{i}$ to a virtual predicate  $pred_{i} \in I$, where $I$ is the pairwise disjoint set of IRIs.  Then, the value of $attr_{i}$ defines $obj_{i}$ as follows:  $obj_{i} := \mu(v(attr_{i})) | \mu(URI(v(attr_{i})))$, 
where $v(attr_{i})$ is the value of $attr_{i}$ and $URI(v(attr_{i}))$ is a URI template populated by
the (API) value of $attr_{i}$ , as the object of a triple can either be a literal or a URI. All URI templates
are defined in the mappings.
Finally, we create a virtual graph $VG_{API, M}$ that consists of triples 
of the form $(subj_{i}, pred_{i}, obj_{i})$. The evaluation of a SPARQL
triple pattern $t$ over a virtual RDF graph on top of an API given the set of mappings $M$, is the following:

\begin{math}
[[t]]_{VG_{API,M}} = \{\mu | dom(\mu) = var(t) \bigwedge \mu(t) \in VG_{API,M}\}.
\end{math}

Notably, even though we mention only Web APIs as data sources, our approach applies uniformly 
to any other non-RDF Web data source as well, such as HTML tables.


\subsection{Methodology}
\label{subsec:methodology}
With the following steps, we can
pose SPARQL queries to non-RDF data sources on-the-fly with the help of the virtual table operator:

\begin{enumerate}
    \item We construct an ontology that models the data of interest.
    \item We create a virtual table operator  for the data source at hand (if it is not available for the kind of data source we want to access (e.g., Twitter API), implementing Algorithm~\ref{vtable}.
    \item We
create the mappings, where the source part 
comprises an \textit{extended-SQL query}, i.e., an SQL query that uses the virtual table operator 
for the selected data source along with the respective parameters. The caching parameter \textit{t} is included optionally as a parameter of the respective virtual tables. 

\item Given the ontology and the mappings, we set up a virtual RDF repository using our extended  \textit{OBDA system} in combination with an
SQL engine that is able to process the extended-SQL queries 
included in the mappings. 
Note that the selected OBDA system 
should 
be (made) ``database-agnostic'' in the sense that it does not require access to the data beforehand. This feature goes beyond the existing
RDB2RDF and OBDA systems, which require that the data to be mapped already reside in a database, to which they connect 
in order to a-priori extract
 meta-data. \cite{ontop-journal,ultrawrap}.
    In our case, we change the OBDA paradigm so that the data is fetched on-the-fly, after a SPARQL query is fired.
    \item Once a  SPARQL query arrives, the OBDA system translates it to SQL. The resulting SQL embeds the virtual table operator(s) involved in the query. 
    \item By the time these operators are invoked as part of the extended-SQL query evaluation, the extended SQL query is evaluated using a system that supports
extended-SQL queries and virtual tables. In our case, this system is MadIS. 
According to the caching parameter \textit{f} that is defined in the mappings, 
MadIS decides whether results will be accessed on-the-fly from the data source
(Step 6a) or cached results will be returned instead (Step 6b). 
\item Eventually, the query result 
returns back to the OBDA system to be presented 
as virtual RDF triples.
\item If applicable, reasoning is applied to the fetched data (e.g., OWL 2 QL  reasoning \cite{motik2009owl} is performed~in~\cite{ontop-journal}). 
\end{enumerate}


\noindent \textbf{Example}. The SQL-extended query described in 
Listing \ref{lst:foursql} includes the virtual table operator
\texttt{foursqr}, which connects to the Foursquare API, retrieves the requested attributes, and populates a virtual table on-the-fly. 
This is not performed a-priori, the virtual table is populated only
when the SQL query is executed. In this way, the most recent version of the data is retrieved, unless the optional parameter \texttt{f} is provided. This parameter defines the length of the window for which cached data can be used. In the case of this example, if the same operator with the same parameters was executed again in less than 10 minutes ago, then the cached data would be returned directly. 

\vspace{-2pt}
\begin{lstlisting}[basicstyle=\ttfamily\scriptsize,frame=tb,caption={Virtual table operator for  Foursquare data},label={lst:foursql}]
 select id, category, name,
 hereNow_count as h, contact  from
 (foursqr key:coffee near:Chicago f:10) ]]
\end{lstlisting}
\vspace{5pt}

Given that our approach 
is generic, we do not associate it with  
a specific mapping language or OBDA system.
Instead, we set 
the specifications such that, once they are met, any RDB2RDF mapping language or system
can implement our approach. Our own implementation is described in
Sections \ref{sec:implementation} and \ref{sec:usecases}.
\section{System Architecture\\ and Implementation}
\label{sec:implementation}


In this section, we  describe the implementation of the above methodology that we implemented
in our system \textsf{Ontop4TheWeb}, which is available opensource as an extension of the system Ontop-spatial\footnote{\url{https://github.com/ConstantB/ontop-spatial}}.
The requirements that the system addresses are the following: 
\begin{enumerate}
    \item To be suitable for data that get updated frequently. 
    \item To be compliant to existing W3C standards for querying RDF data (either materialised or virtual), i.e., the W3C standard SPARQL query language should be supported. Applications that build on top of SPARQL should be able to use this system regardless of the underlying implementation and/or the format of the original data sources.
    \item To support different data formats,
    but to represent them, on the high level, using a uniform schema. 
    \item To represent data as virtual RDF terms, thus 
    saving users from converting Web data 
    via a set of tools specialized for parsing, converting and storing data
    as RDF triples. 
\end{enumerate}

The design choices that address these requirements are the following:
\begin{enumerate}
    \item We create a virtual table operator for every \textbf{kind} of data source (not for every data source)  that we want to represent as virtual RDF graph. These operators are embedded in SQL queries and, once invoked , these operators connect to the original data sources and return the data in tabular format (virtual tables). 
    \item Virtual table operators include a caching mechanism, using the same data they were previously retrieved in a previous execution, for a time window $w$. The length of this window is given as a parameter so that it can be adjusted according to the requirements of a specific use (e.g., Velocity is typically different for every data source).
    \item Standard SPARQL queries are provided as input to the system. \item OWL2 QL ontologies are also provided as input to the system to model the data regardless of the original format of the data source. The W3C standard mapping language R2RML is used to encode how the data from the virtual tables can be mapped to virtual RDF terms. For the first time, R2RML mappings include SQL queries with embedded virtual table operators, thus connecting data that are not materialised in a DBMS to virtual RDF terms. 
    \item We extend the OBDA paradigm in order to connect to virtual relational data and, using ontologies and mappings, create virtual RDF graphs on top of them. By enabling on-the-fly SPARQL-to-SQL translation, this data can be queried using SPARQL, in the same way that one could query the data as if it had been converted into RDF and stored in triple stores.
\end{enumerate}

The architecture of \textsf{Ontop4theWeb}  is shown in Figure \ref{fig:arch}.
The system consists of the following components:

\begin{figure}[t]
\begin{center}
   	\includegraphics[width=0.52\textwidth]{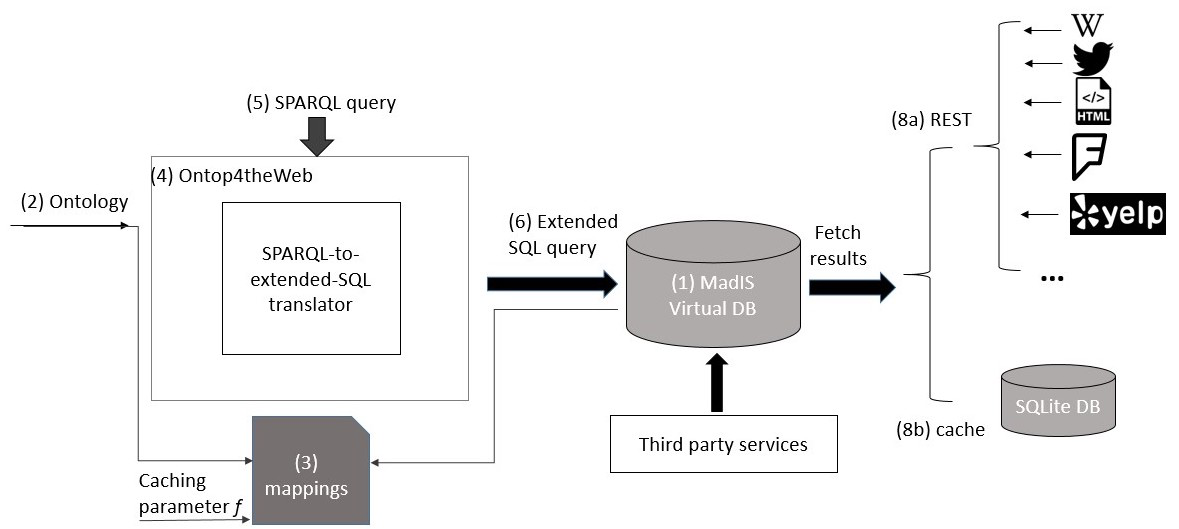}
   \vspace{-5pt}	
   \caption{System architecture for \textsf{Ontop4theWeb}}.
   \vspace{-5pt}
\label{fig:arch}
\end{center}
\end{figure}

$\bullet$ The back-end is based on \textit{MadIS}\footnote{\url{http://madgik.github.io/madis}} \cite{madis}, a relational database system that relies on \textit{SQLite}\footnote{\url{http://www.sqlite.org}}, but extends it via the Python wrapper \textit{APSW}\footnote{\url{https://github.com/rogerbinns/apsw}}. The SQLite database can be extended with user-defined operators that can be used as row, aggregate, or virtual table operators. To this end, MadIS exploits the APSW SQLite wrapper, which provides an 
    interface for implementing such operators in an extensible way through Python.
    Using APSW, we define our own operators to create virtual tables and populate them with data retrieved from the Web. To query the retrieved data, we use \textit{MadQL},
    the MadIS implementation of the extended-SQL language described in Section \ref{sec:approach}, which
    contains the virtual table operators. 
    Instead of using MadIS, we could implement
    the same virtual table operators in C, extending SQLite directly, but this would be less user-friendly and re-usable than the plug-and-play MadIS Python operators; it would also undermine the modularity and extensibility of \textsf{Ontop4theWeb}.
    
$\bullet$ Third party applications are external \textit{micro-services} that can be invoked by a virtual table operator in MadIS. For example, 
    a virtual table operator is able to identify the sentiment of a tweet by calling a micro-service that implements a Sentiment Analysis classifier
    (see Section \ref{subsec:twitter}). 
This feature enables 
\textsf{Ontop4theWeb} to perform data analysis tasks without facing any 
compatibility issues between the virtual table operator and any data analysis software: the server can be written in any 
language or platform, but the client can still use it as a service.

$\bullet$ \textit{Ontop}\footnote{\url{https://github.com/ontop/ontop}} \cite{ontop}
is a state-of-the-art, open-source OBDA system that supports R2RML and its native mapping language. 
We extended the MadIS JDBC connector so that it complies with  Ontop, while Ontop was extended to use MadIS as a back-end.
        The latter modification is the most significant one,
        enabling Ontop to operate in a ``database-agnostic'' manner that supports non-materialized databases and relies on MadIS as back-end.
        The reason is that Ontop, like all other OBDA systems, connects only to
        populated and materialized databases, using their data for optimization, \emph{before} querying them. 
        Instead, \textsf{Ontop4theWeb} retrieves the data to be queried only \emph{after} the user fires a query, creating a virtual table on-the-fly. As a result, no prior knowledge of the data can be used.
  
\section{Practical Scenarios}
\label{sec:usecases}

We now showcase how we can pose SPARQL queries on data coming from WebTables or REST APIs using \textsf{Ontop4theWeb}.

\subsection{HTML tables use case}
\label{subsec:webtables}

HTML tables constitute one of the most common tabular formats for publishing data on the Web.
A lot of research activities and applications
have focused on retrieving, mining, annotating, and semantically-enriching information available in WebTables \cite{WebTables}. As an example, consider a semantic-based recommendation engine  that tries to address the cold-start problem for new users. To make meaningful suggestions for users with empty profile and no history, it uses the American Film Institute list of the 100 best movies from Wikipedia\footnote{\url{http://en.wikipedia.org/wiki/AFI\%27s\_100\_Years...100\_Movies}} in combination with the latest list of user reviews from Rotten Tomatoes\footnote{\url{http://www.rottentomatoes.com/top/bestofrt/}}, as shown in Figure \ref{fig:films}. This is expressed 
with the SPARQL query described in Listing \ref{lst:WebTable}.

\begin{lstlisting}[basicstyle=\ttfamily\scriptsize,frame=tb,caption={Querying WebTables using SPARQL},label={lst:WebTable}]
PREFIX wiki: <http://en.wikipedia.org/movies/ontology#>
PREFIX r: <http://www.rottentomatoes.com/top/bestofrt/>
						
select distinct  ?title ?rrank ?wrank
where { 
?s r:title ?title .
?s2 wiki:title ?title . 
?s r:rank ?rrank .
?s2 wiki:rank ?wrank  }
\end{lstlisting}

The SPARQL query provided  in Listing \ref{lst:WebTable} retrieves the titles of movies that are included in both tables and  the respective ranks. This is performed by
executing a join on the ``title" column of both tables. We now explain  how we can accommodate this application using \textsf{Ontop4theWeb} to query data contained in HTML tables based on ontologies and mappings. 

First, we use the virtual table operator \texttt{WebTable}, extending the respective MadIS operator. This operator
creates a virtual table and populates it with data contained in
the HTML table that is given as input so that this data can be queried
using MadQL queries. These queries can then be embedded in mappings 
as a data source, creating virtual RDF graphs.


The mappings provided in Listing \ref{lst:WebTables} describe how the information contained in these tables
is translated into RDF terms. From the Rotten Tomatoes WebTable, we retrieve the rank number of films according to reviews  along with the title of the film. 
From the Wikipedia WebTable, we retrieve the title, the ranks for years 1998 and 2007 and the release date. The aim of this task is to compare and combine two different sources of information (Wikipedia and Rotten Tomatoes) based on the ranks of movies. 
To retrieve this
information, we use the \texttt{WebTable} virtual table operator that parses an HTML table and returns the results as a virtual
table. The MadQL query that uses this operator can be seen in both mappings. Its first argument is the HTML page that contains
the respective WebTable, while the second one is the index of the WebTable in the page. In our example, we want the 3$^{rd}$ HTML table that appears in the Rotten Tomatoes
page and the 2$^{nd}$ one that appears in the respective Wikipedia page. 

Note that the Rotten Tomatoes website includes the release date of every film in parenthesis next to the film title, while
the Wikipedia table provides it in a separate column. Since we want to join the two WebTables  on the ``Title'' field, 
we 
align this attribute so that it has the same format in both tables. To achieve this,  we concatenate 
the columns ``Title'' and ``Release year" of the Wikipedia table so that the format of the resulting title
is
exactly the same with the one in the Rotten Tomatoes WebTable. 

\begin{lstlisting}[basicstyle=\ttfamily\scriptsize,frame=tb,caption={Mapping for WebTable data},label={lst:WebTables}]
[MappingDeclaration] @collection [[

mappingId WebTable_rotten_tomatoes
target rot:{rank} rot:rank {rank};
       rot:title {Title};
       rot:reviews  {reviews}^^xsd:int; 
       rot:rating {RatingTomatometer}^^xsd:int .
source select  rid as rank,
       "No. of Reviews" as reviews,
       Title, RatingTomatometer from 
WebTable('http://www.rottentomatoes.com/
top/bestofrt/',3)

mappingId WebTable_wikipedia
target wiki:{rid}  wiki:title {Title}; 
       wiki:rank98 {rank98}^^xsd:int ;
       wiki:rank {rank} .
source  select rid, rank,Title from 
       (select rid,Film||" ("||"Release year"||")"
       as Title,"2007 rank" as rank from WebTable(
'http://en.wikipedia.org/wiki/AFI\%27s\_100
\_Years...100\_Movies',2))]]
\end{lstlisting}

\subsection{Twitter Use Case}
\label{subsec:twitter}
Twitter is a popular social network whose popularity is increasing to the extent that many people use it as a news stream \cite{twitter}. Collecting its data is important for many  academic and commercial activities to perform data mining, integration, and analysis tasks \cite{twitteranalysis}. 
Twitter data sources have the following characteristics (Velocity)σ: (i) They get frequently updated (about 8,000 tweets are posted per second and around 700M are posted per day\footnote{\url{http://www.internetlivestats.com/twitter-statistics/}}),
(ii) They are more important when they are fresh - the primary use of Twitter is to  find out information about \textit{what is happening now}. (iii) They are frequently
used by data scientists as input datasets to data analysis and data mining tasks (e.g., 
sentiment analysis \cite{sentanalysis}).

Typically, users write crawlers to retrieve Twitter data and store it in files or in a database. Since the Twitter API has a limit
of 100 tweets per request, 
the crawlers 
perform multiple requests and accumulate
data over a large period of time. 
Let us now imagine a semantic-based application that tracks user-generated content about
active academia events,
collecting the latest relevant tweets 
and processing them
with a sentiment analysis service that identifies their polarity.
For example, it uses the  SPARQL query described in Listing \ref{lst:twittersparql} to retrieve positive tweets about EDBT 2020.

\begin{lstlisting}[basicstyle=\ttfamily\scriptsize,frame=tb,caption={Querying Twitter using SPARQL},label={lst:twittersparql}]
select distinct ?s  
where { 
?s twitter:tweetsAbout 
<https://diku-dk.github.io/edbticdt2020> . 
?s twitter:sentiment "positive"}
\end{lstlisting}


Traditionally, this SPARQL query would be answered through
the following steps: (i) retrieve the relevant Twitter data, (ii) transform it into RDF, and (iii) store it in a RDF store. Alternatively, one would store the data in a database, using an OBDA system to query it with mappings. The sentiment analysis task would be performed as a pre- or a post- processing step.

In contrast, using \textsf{Ontop4theWeb} requires less steps for
answering this query.  After a 
SPARQL query like the one described above is fired, a virtual table is created, containing information about every tweet along with its \texttt{sentiment},
i.e., whether its sentiment is positive or negative. Then,  this information gets mapped into virtual RDF terms.
To 
this end, we implemented a virtual table operator that (i) searches data using the Twitter REST API,  (ii) uses a 
binary 
classifier to identify whether it is
positive of negative, and (iii)  populates a virtual table with the results. Its data are then accessed via MadQL queries that can be incorporated in mappings so that virtual RDF triples can be
produced on-the-fly. An exemplary mapping appears in Listing~\ref{lst:twittermap}.

\begin{lstlisting}[basicstyle=\ttfamily\scriptsize,frame=tb,caption={Mappings for Twitter data},label={lst:twittermap}]
[MappingDeclaration] @collection [[
mappingId       twitter_mapping
target          twitter:{username} twitter:tweetsAbout 
                <https://diku-dk.github.io/edbticdt2020>;
                twitter:sentiment {sentiment}.
source          select distinct id, sentiment 
                from (twitterapi key:edbt2020) ]]
\end{lstlisting}

The source part of this mapping 
contains a MadQL query that uses
the virtual table operator named {\small\texttt{twitterapi}}. This virtual table operator takes as input a search keyword, which in our example is \emph{edbt2020}. The result of this query is the creation
of a virtual table with information about  tweets for EDBT 2020. Note that the attribute {\small \texttt{sentiment}} is not part of the data
retrieved from Twitter API, but is 
derived from
the sentiment analysis classifier that is used
internally, in the {\small\texttt{twitterapi}} virtual table operator.

In this context, the SPARQL query in Listing \ref{lst:twittersparql} is translated into the SQL query in Listing \ref{lst:sqlquery}.

\begin{lstlisting}[basicstyle=\ttfamily\scriptsize,frame=tb,caption={SQL query for the virtual table of Twitter},label={lst:sqlquery}]

SELECT * FROM ( SELECT DISTINCT 
1 AS "sQuestType", NULL AS "sLang", ('http://twitter.com/' ||
REPLACE(REPLACE(REPLACE(REPLACE(REPLACE(REPLACE(REPLACE(REPLACE
(REPLACE(REPLACE(REPLACE(REPLACE(REPLACE(REPLACE(REPLACE(REPLACE
(REPLACE(REPLACE(REPLACE(CAST(QVIEW1.id AS CHAR),' ', '%20'),
   '!', '%21'),...)) AS "s" FROM 
(select distinct id, 
sentiment from (twitterapi key:edbt2020)) QVIEW1,
(select distinct id,
sentiment from (twitterapi key:edbt2020)) QVIEW2
WHERE  QVIEW1.id IS NOT NULL AND
(QVIEW1.id = QVIEW2.id) AND
(QVIEW2.sentiment = 'positive')) SUB_QVIEW;
\end{lstlisting}

This query 
contains the virtual table operator \texttt{twitterapi}
that creates a virtual table. The columns 
\texttt{id} and \texttt{sentiment}
of this table 
populate a view that is created on-the-fly by the OBDA system. In traditional 
OBDA systems, the views are constructed on-the-fly  from existing, materialized tables (or other views). In \textsf{Ontop4theWeb}, this table does not exist, but
is created and populated on-the-fly, \textit{after} the SPARQL query is fired and translated
into MadQL. The MadQL query will create and populate the table, but this procedure is
completely invisible to the user: 
exactly the same SPARQL query would be used even if the data did not come from a REST API, but was stored in a database, or a triple store. 

To classify the tweet according to its polarity, we 
employed an open-source
sentiment classifier for Twitter\footnote{\url{https://github.com/dkakkar/Twitter-Sentiment-Classifier}}, which uses an SVM model that is already trained with the following datasets: (i) The Stanford Sentiment140 dataset\footnote{\url{http://cs.stanford.edu/people/alecmgo/trainingandtestdata.zip}}, (ii)
the Polarity Dataset (v2.0)\footnote{\url{http://www.cs.cornell.edu/people/pabo/movie-review-data/}},
and (iii) a dataset from the
University of Michigan\footnote{\url{https://inclass.kaggle.com/c/si650winter11}} that contains 7,086 sentences extracted from various social media.

We have modified this classifier so that it follows a client-server model, where the
server and the client communicate through a socket. 
In this way, we avoid incorporating the whole classifier
into the virtual table operator and save the cost of 
loading the classifier every time
the virtual table operator is invoked. 
When the server starts, it loads
the classifier and waits for connection. The client part is incorporated into the \texttt{twitterapi}
virtual  table operator and sends every tweet of
the results to the server for classification 
through
a socket. The server performs sentiment analysis and returns whether the
tweet is positive or negative. The result is returned as an additional column of the
produced virtual table, called \texttt{sentiment}.

\subsection{Foursquare Use Case} 

Foursquare\footnote{\url{https://foursquare.com}} is a mobile application that offers location-based  search for venues with multiple criteria (e.g., nearby restaurants 
ranked by rating or distance). The descriptions of these venues are enriched with user reviews and ratings, thus facilitating location recommendations. Foursquare also allows users to share their location with their friends, informs them how many other users are simultaneously at the same location, and alerts them when many people check in at the same time in a place nearby.

Having around 55 million monthly users and a platform that contains
crowd-sourced information for 105 million venues worldwide (according to its website), Foursquare has become a useful
data source for various applications. Developers can access its API\footnote{\url{https://developer.foursquare.com/}} and get part of this information for free (e.g., venue description, location, rating, check-ins), while
more data is available on charge. Foursquare has approximately 40,000 registered
developers using its API\footnote{\url{https://en.wikipedia.org/wiki/Foursquare\#Foursquare_API}}.
As an example of applications built on top of Foursquare, consider the ``Mr Jitters" app, which uses Foursquare data to find the best coffee places nearby\footnote{\url{https://developer.foursquare.com/docs/sample-apps}}.

Semantic Web agents could also exploit this valuable data source.
Imagine a semantic-web alternative to ``Mr Jitters" 
that 
uses Foursquare venues as RDF so as to 
interlink them with datasets from the linked open data cloud (e.g., LinkedGeodata). 
Supposing that it searches 
information about coffee places
in Chicago, it would pose the SPARQL query described in Listing \ref{lst:foursparql}.

\begin{lstlisting}[basicstyle=\ttfamily\scriptsize,frame=tb,caption={Querying Foursquare using SPARQL},label={lst:foursparql}]
select ?venue ?checkins 
where {
?venue four:name ; 
four:hereNow ?checkins;
four:category "Coffee";
four:near "Chicago"}
\end{lstlisting}

\textsf{Ontop4theWeb} can be used to map the free Foursquare data
to virtual RDF graphs and perform this query on top of them. First, we create an ontology
that describes all venue categories that appear in the Foursquare venue category taxonomy\footnote{\url{https://developer.Foursquare.com/docs/resources/categories}}.
The resulting ontology\footnote{\url{http://pyravlos-vm5.di.uoa.gr/foursquare.owl}} 
contains a rich hierarchy of 961 classes that represent venue categories,
enabling reasoning over it.

Next, we implement a virtual table operator, called \texttt{foursqr}, which receives as input some keywords for searching 
venues and returns as output a list of venues. The 
operator is implemented 
as a Python MadIS virtual table operator, which internally uses a Python library
for the Foursquare API\footnote{\url{https://github.com/mLewisLogic/foursquare.git}}.
When the Foursquare virtual table operator is invoked, it accesses the Foursquare API
with the input parameters as arguments, and the result is presented as a virtual table that in turn gets mapped into RDF terms using the  mapping described in Listing \ref{lst:fourmap}.

\begin{lstlisting}[basicstyle=\ttfamily\scriptsize,frame=tb,caption={Mapping for Foursquare data},label={lst:fourmap}]

[PrefixDeclaration] four: http://foursquare.com/
[MappingDeclaration] @collection [[
 mappingId foursquare_mapping
 target    four:{id} four:hasID {id} ;
           four:name {name} ;
           four:hereNow {h}^^xsd:integer;
           four:category four:{category};
           four:near "Chicago"  .
 source    select id, category, name,
           hereNow_count as h, contact 
           from (foursqr key:coffee near:Chicago) ]]
\end{lstlisting}
In this mapping,
we want to retrieve coffee places in Chicago. The 
\texttt{foursqr} operator used in the MadQL query of the mapping takes
the respective parameters as input. It generates a virtual table populated with information about coffee places in Chicago. The target
part of the mapping encodes how these attributes are translated into
RDF terms according to the Foursquare Ontology. 
\section{Experimental evaluation}
\label{sec:evaluation}



\subsection{Experimental setup}


\textbf{Execution environment}. All experiments run on a PC with Intel Core\texttrademark\\ 2 Quad CPU Q9650 at 3.00GHz, 8GB RAM and Ubuntu 14.04. In all experiments, we measure the query execution time, which includes a full iteration over the result set. We repeat every execution 3 times and consider the average running time. We execute all experiments in both cold and warm cache. In warm cache, we execute a query once before all executions of the same query that we measure. 
In cold cache, we configure all virtual tables that are involved so that they do not use the caching mechanism described in Section \ref{sec:approach} (i.e., we set 
a negative value to the \texttt{rate} parameter of the virtual table operator in the mappings). These two configurations allow for measuring the impact of the caching mechanism 
on the query execution time. 

\vspace{2pt}
\noindent
\textbf{Data sources and queries}. We query data from WebTables and REST APIs, using queries that are similar to the ones described in Section \ref{sec:usecases}. More specifically, we pose queries for tweets that contain the \texttt{EDBT2020} hashtag, retrieving also the sentiment for each tweet. We look for coffee places in Chicago from Foursquare and we join two HTML tables with films, one from Wikipedia and one from Rotten Tomatoes. The mappings and part of the queries that we used are explained in Section \ref{sec:usecases}. For each data source,
we begin with a query that involves a single triple pattern and then, we increment the number of triple patterns to increase
the complexity of the query. 

To evaluate the scalability of \textsf{Ontop4theWeb}, we use synthetic WebTables. 
We employed an original Wikipedia table  about Italian election opinion polls\footnote{\url{https://en.wikipedia.org/wiki/Opinion\_polling\_for\_the\_Italian\_general\_election,\_2018}} as a template, which  we multiplied  so that we can execute queries for tables with 10, 100, 1,000, 10,000, and 100,000 rows. Then, we posed the same queries over these tables in order to measure the scalability of \textsf{Ontop4theWeb}. 

\subsection{Experimental Results}


\begin{figure}[t!]
\begin{center}
   	\includegraphics[scale=0.6]{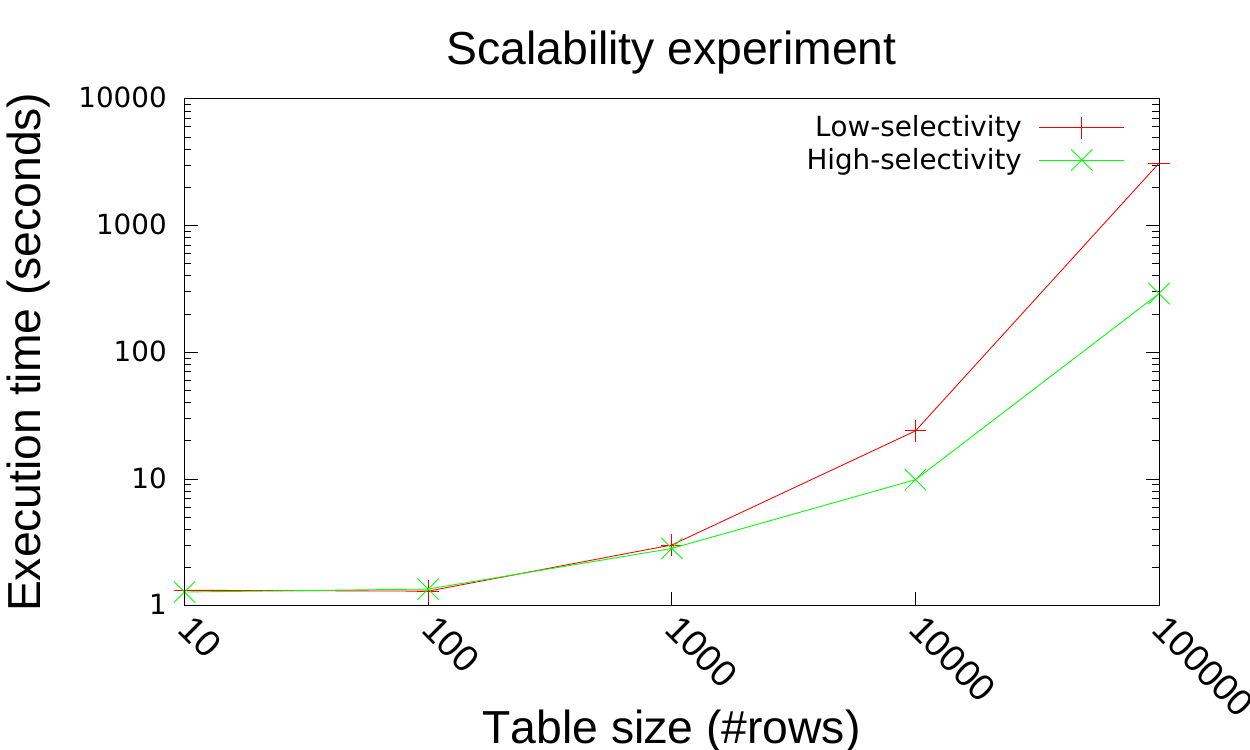}
   \caption{Query execution time as dataset size increases.}
\label{fig:expscal}
\end{center}
\end{figure}

\begin{figure*}[t]
 \setlength{\abovecaptionskip}{0.0pt}
 \includegraphics[width=.45\textwidth]{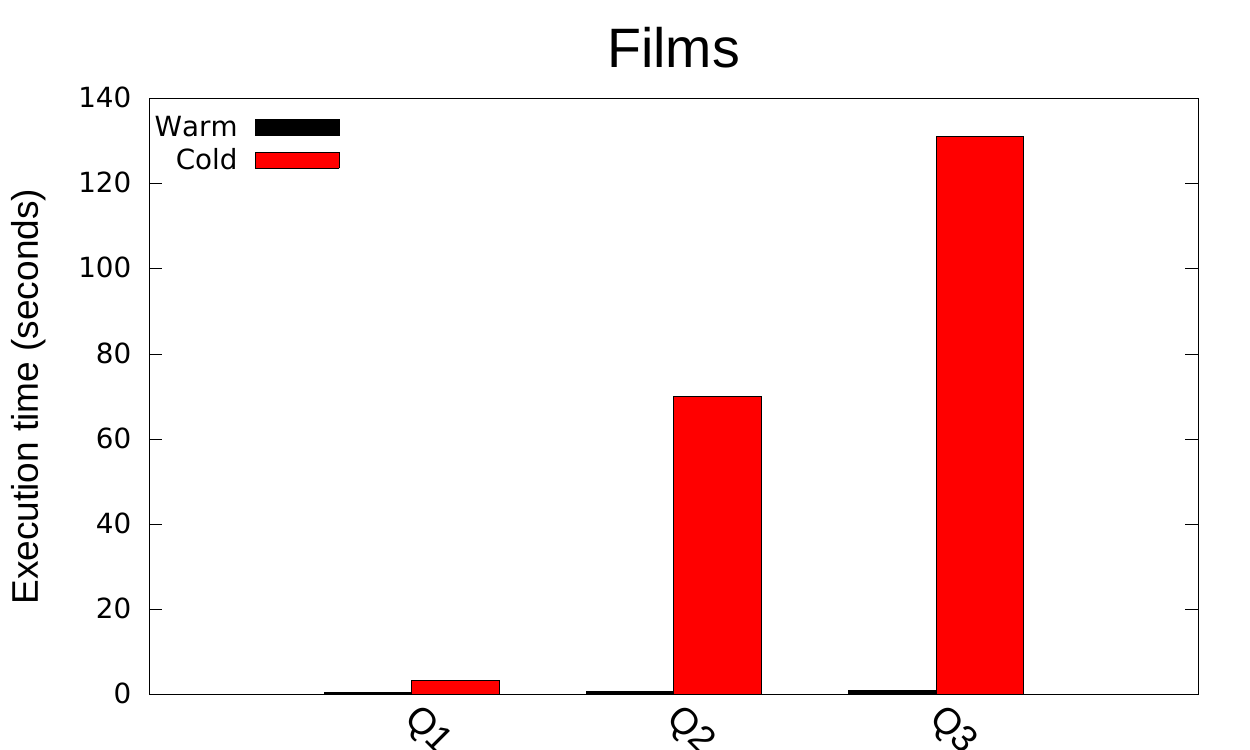}
\includegraphics[width=.45\textwidth]{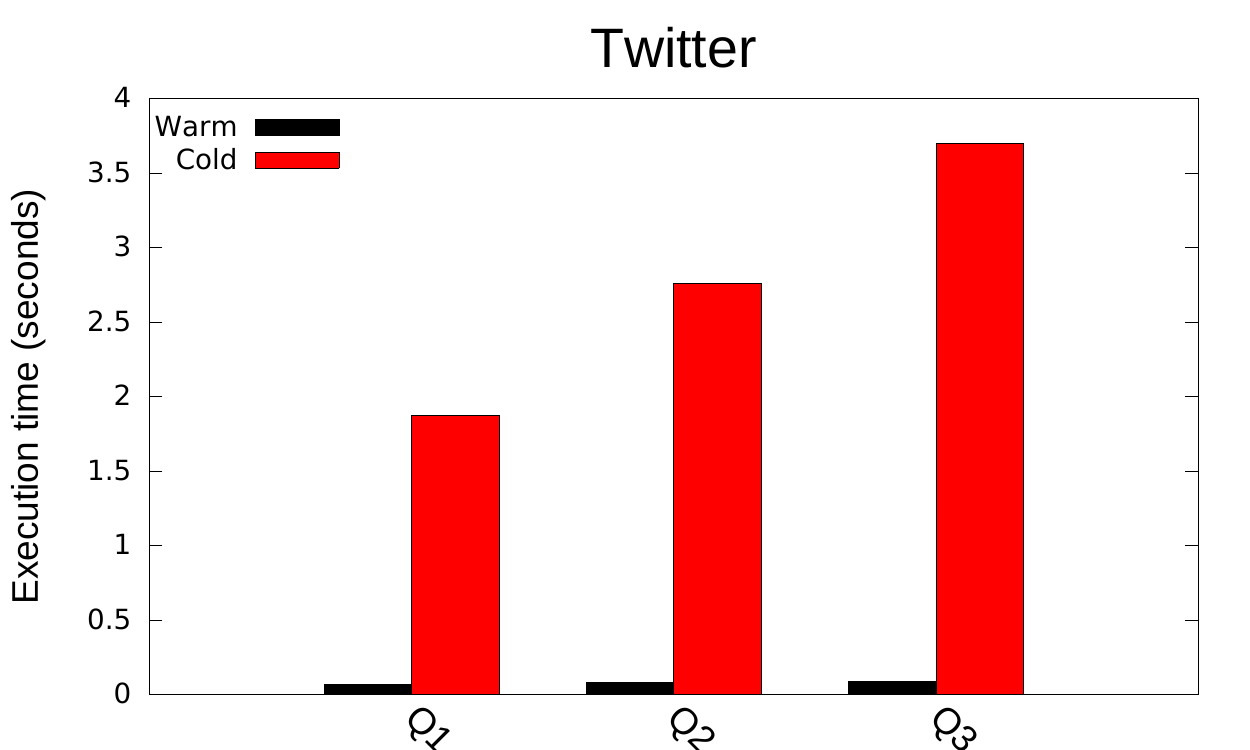}
\includegraphics[width=.45\textwidth]{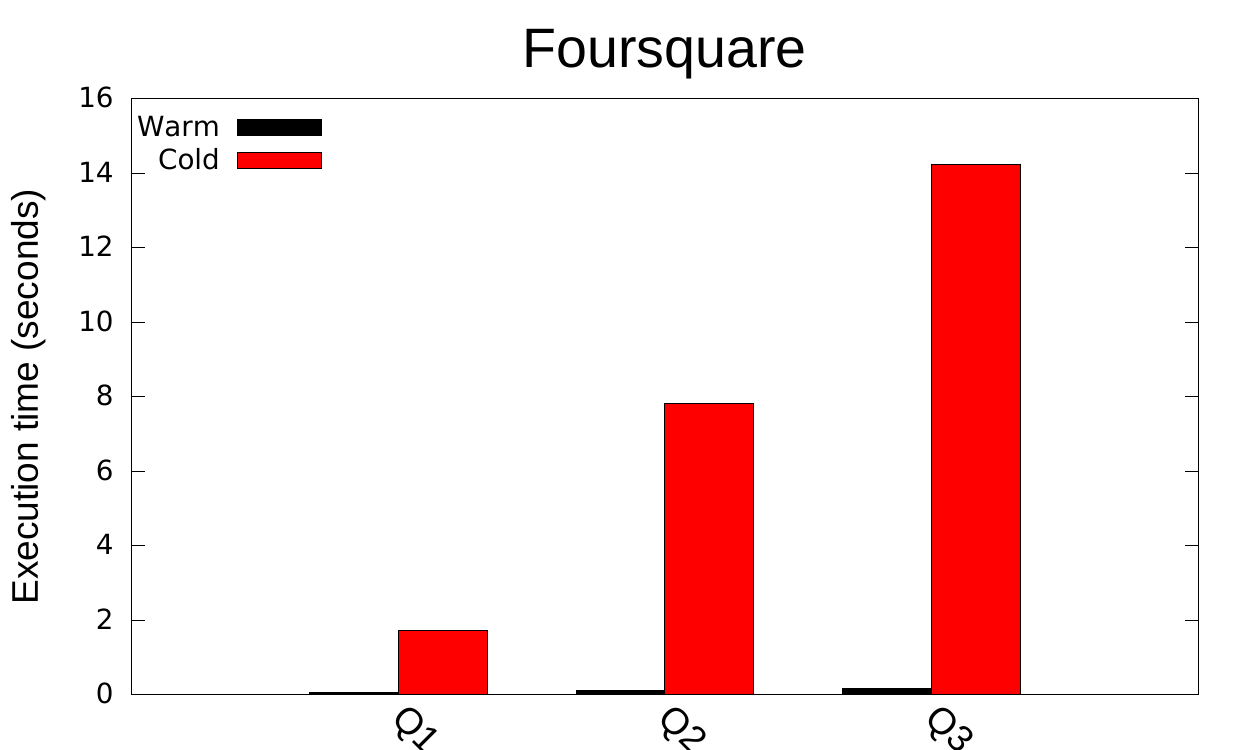}
\caption{Execution times for real workload queries.}
\label{fig:expreal}
 \end{figure*}
 
\textbf{Real workload}. The query execution times of the real workload
experiments in both cold and warm cache 
are presented
in Figure \ref{fig:expreal}. The label of each query 
is suffixed by the number of triple patterns it incorporates (e.g., Q2 indicates two triple patterns).

We observe that 
the execution times in warm cache are at least an order of magnitude lower than in cold cache and that they remain stable, regardless of the query complexity.
In contrast, 
as the number of triple patterns in the queries increases, the execution time in cold cash increases considerably.
This happens because more triple patterns yield more joins in the translated SQL query. When these joins produce more intermediate results, instead of filtering them down, they introduce additional cost in the evaluation. 
In other words, 
we add triple patterns to retrieve more information, rather than to pose restrictions.
The main reason 
is that the data is not materialized in the database
and, thus, the OBDA system is not aware of database constraints, or other hints
that could accelerate SQL translation and execution, as described in \cite{ontop}. 

Note also that all films queries use two data sources, joining the Webtables described in Section \ref{subsec:webtables} to retrieve the movies that are common between the two tables. This involves a higher cost than the Twitter and Foursquare use cases, 
as the query execution time also includes the time to parse the HTML table(s). 

\vspace{2pt}
\noindent
\textbf{Synthetic workload}. 
The goal of our scalability analysis is to assess the maximum size 
of input data 
that we can query efficiently.
We used two queries with two triple patterns posed against the synthetic
WebTables of varying size. The first query, which is provided in Listing \ref{lst:weblow}, is not selective, returning as many results as the rows of the table. The second query, which is described in Listing \ref{lst:webhigh} is very selective, returning two results at all cases. 

\begin{lstlisting}[basicstyle=\ttfamily\small,frame=tb,caption={Query of low selectivity for WebTables},label={lst:weblow}]
select distinct ?s1 ?d ?l  
where {
?s1 :date ?d .  
?s1 :lead ?l .}
\end{lstlisting}

\begin{lstlisting}[basicstyle=\ttfamily\small,frame=tb,caption={Query of high selectivity for WebTables},label={lst:webhigh}]
select distinct ?s1 ?d   
where {
?s1 :date ?d 
?s1 :lead \"1.5\
"^^<http://www.w3.org/2001/XMLSchema#float> . }
\end{lstlisting}

The outcomes of the scalability test appear in Figure \ref{fig:expscal}. We observe 
that as the number of rows in a WebTable increases, the query execution time increases superlinearly, but the extent of this increase depends heavily on the selectivity of the query.
We observe, though, 
that \textsf{Ontop4theWeb} can process queries against WebTables with up to 100,000 rows
within minutes, when the selectivity is high.

\subsection{Comparison with the state-of-the-art}

\subsubsection{Qualitative comparison}
We now compare \textsf{Ontop4theWeb} with the \textit{SERVICE-to-API} system \cite{eswc18}.\footnote{We also attempted to compare \textsf{Ontop4theWeb} with the work described in \cite{ldow18}, but we could not build an instance of their platform, following the online instructions.} Recall that its goal is to enrich RDF data with data from external sources, such as REST APIs. Thus, its query language requires at least one triple pattern that is evaluated in the RDF repository and its variables are bounded to values that populate their URI templates. Every variable binding actually yields a separate API call. A cache mechanism aims to minimize the API calls.

An example of its query language is presented in  Listing \ref{lst:yelp3a}. The value of keyword $SERVICE$ creates a URI template for each one of the values bound to the variable $l$, which is used in the query's triple pattern. In this case, a call to the Yelp API is produced for each binding of the variable $l$, returning a JSON file. This JSON file is parsed according to the JSON pattern included in the query, which bounds the variables $i$, $name$ and $rating$ to the values of the respective attributes of the JSON file.

\begin{lstlisting}[basicstyle=\ttfamily\scriptsize,frame=tb,caption={SERVICE-to-API query, equivalent to SPARQL query Q1 in Listing \ref{lst:yelp}},label={lst:yelp3a}]
SELECT   ?i ?name ?rating WHERE {
?x <http://www.w3.org/2000/01/rdf-schema#label> ?l . 
SERVICE <https://api.yelp.com/v3/businesses/{l}>{
( $.[\"id\"], $.[\"name\"],
$.[\"rating\"]) AS (?i, ?name, ?rating)}}
\end{lstlisting}

In this context, there are two major \textit{qualitative} differences between \textsf{Ontop4theWeb} and SERVICE-to-API \cite{eswc18}:
\begin{enumerate}
    \item The query language. For SERVICE-to-API, the JSON attributes are directly bound to variables by parsing the JSON response, as instructed by the JSON patterns included in the query. As a result, the users need to know the documentation of the API in order to identify the information they need. Only in this way are they able to combine API data with the RDF data in the triplestore, formulating accurate queries that extend SPARQL with JSON patterns \cite{eswc18}. In contrast, \textsf{Ontop4theWeb} creates virtual semantic graphs on top of APIs using mappings, thus allowing users to pose standard SPARQL queries as if the contents of the APIs were transformed into RDF. The trade-off for not having to convert, materialise and store the data into an RDF store is the use of mappings. For any virtual \textsf{Ontop4theWeb}  RDF repository, a mapping file should  be provided. On the one hand, writing the mappings can be an overhead. However this approach has the following advantages: (i) mappings need to be written once unless the schema changes, (ii) the mapping language R2RML is W3C standard, as well as the SPARQL query language. This ensures compatibility with applications built on top of SPARQL, (iii) materialisation is not avoided in  SERVICE-to-API, part of the data needs to be stored in a triple store.
    \item Every API call in \textsf{Ontop4theWeb} retrieves an entire virtual table, which is mapped to a virtual RDF graph. In contrast, SERVICE-to-API merely retrieves one entry of this table per API call, which has a significant impact on time efficiency, as explained in the following experiments.
    However, both systems use a cachine mechanism
\end{enumerate}

 \begin{figure}
 \setlength{\abovecaptionskip}{0.0pt}
\includegraphics[width=.4\textwidth]{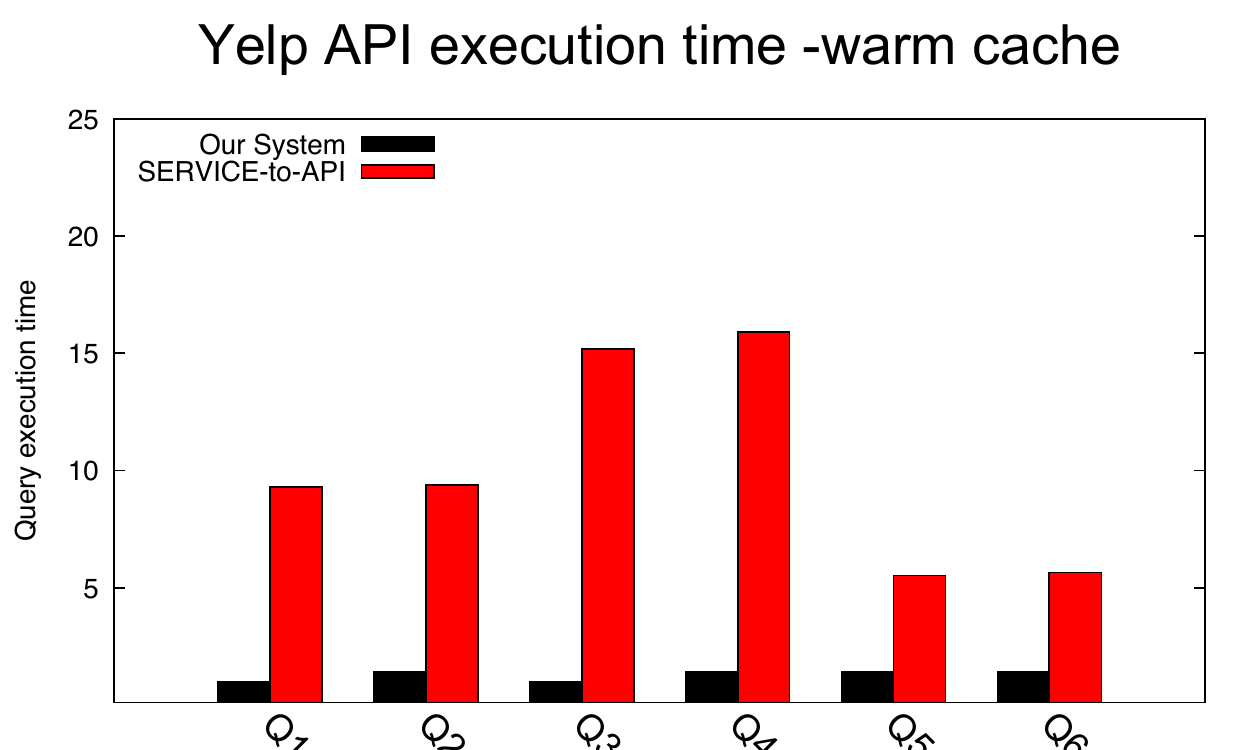}
\includegraphics[width=.4\textwidth]{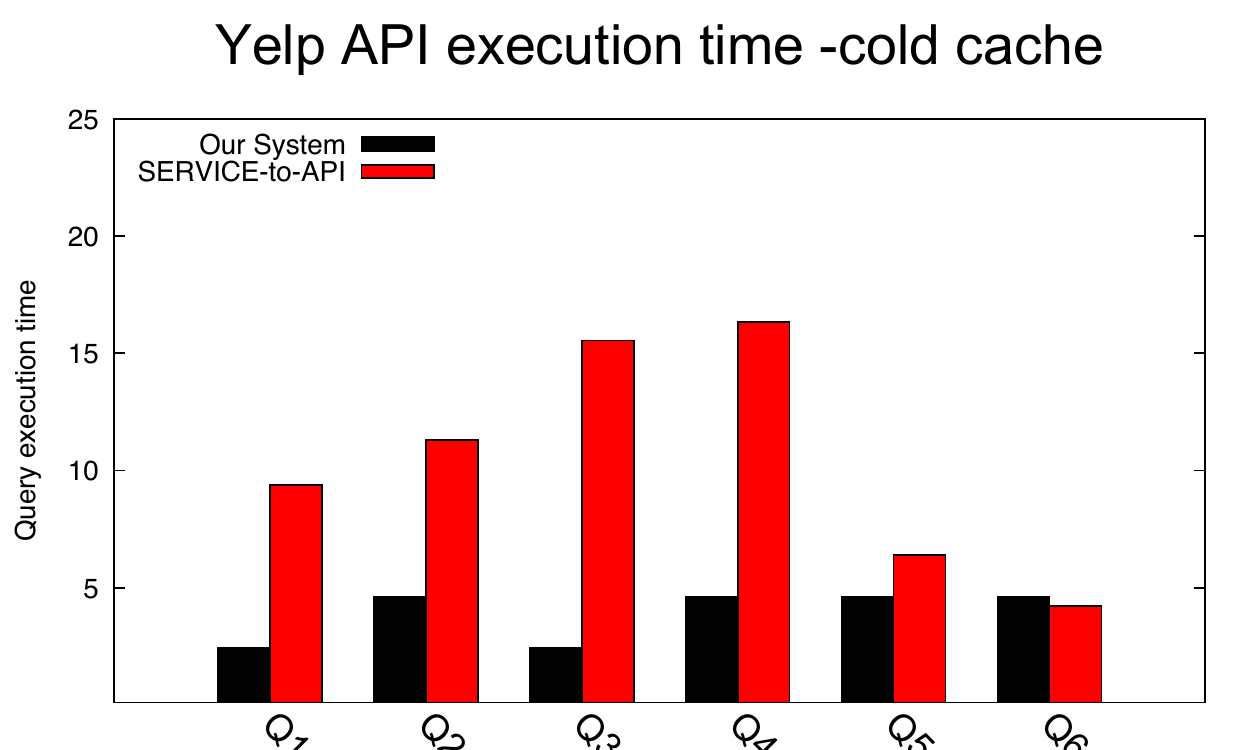}
\caption{Execution times  for Yelp queries in warm and cold cache.}
\label{fig:expyelp}
 \end{figure}
 
 \begin{figure}
 \setlength{\abovecaptionskip}{0.0pt}
\includegraphics[width=.4\textwidth]{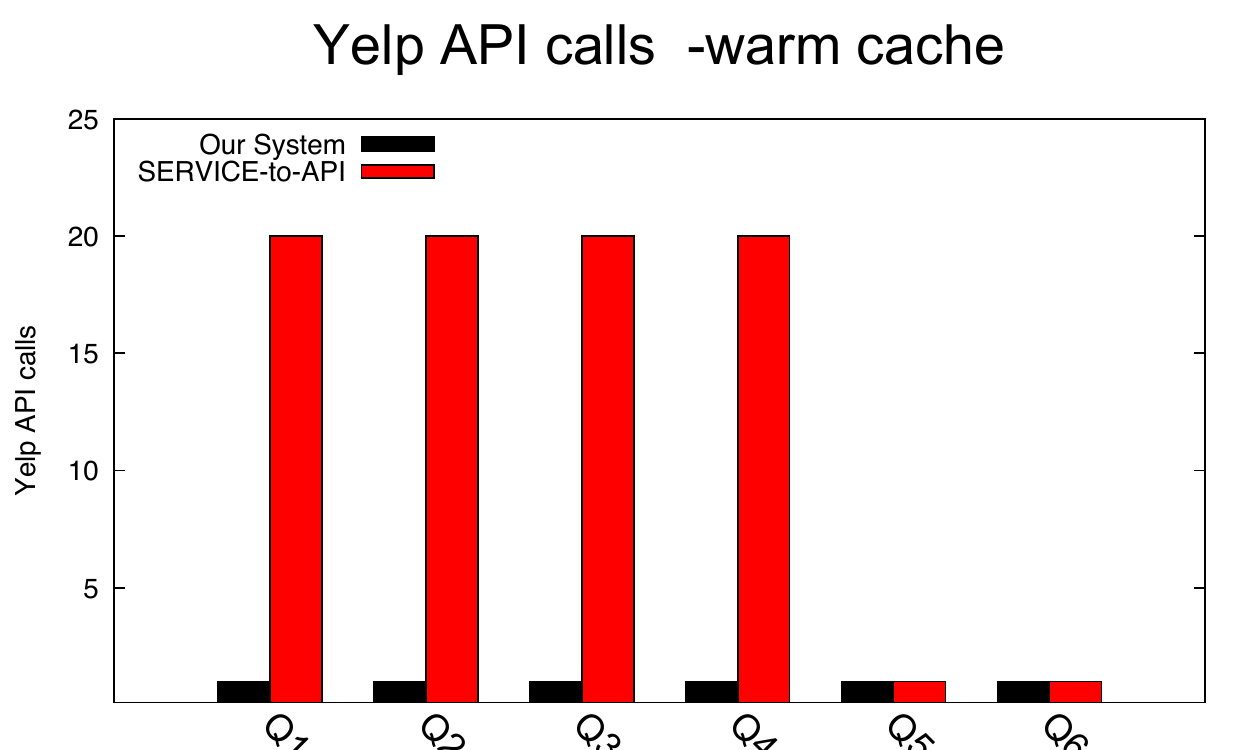}
\includegraphics[width=.4\textwidth]{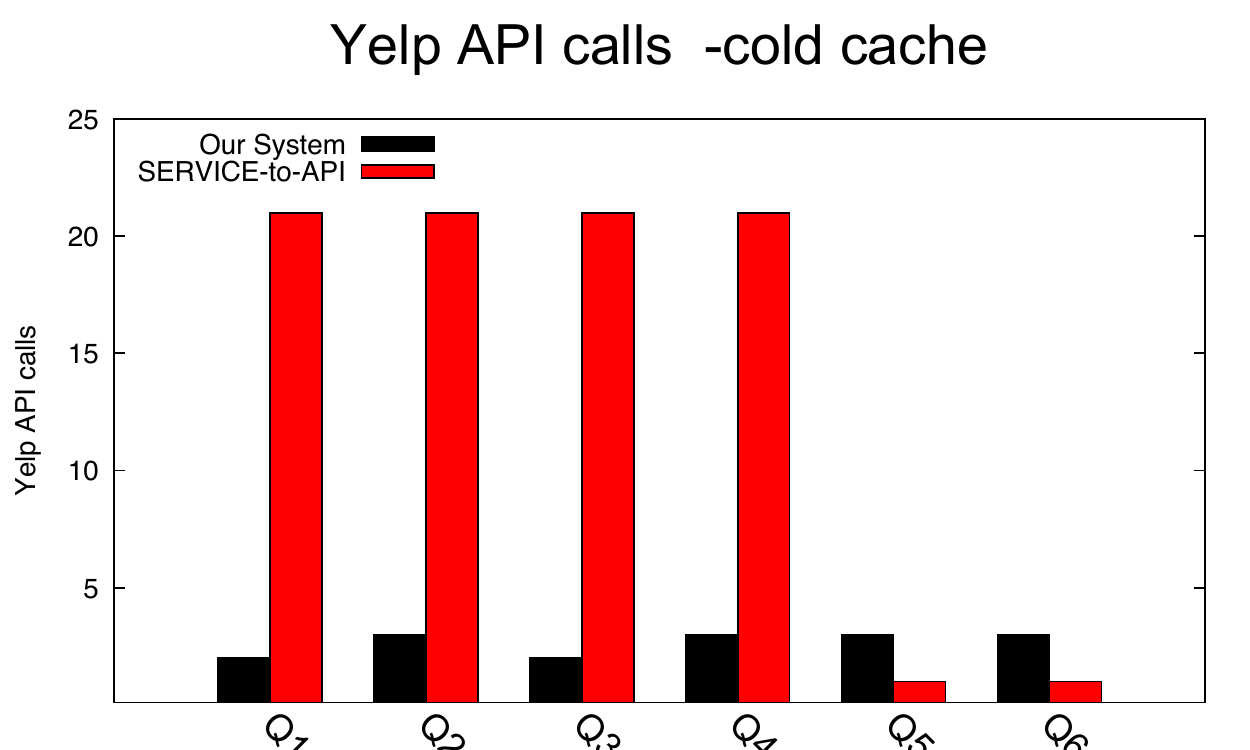}
\caption{API calls for Yelp queries in warm and cold cache.}
\label{fig:expyelpcalls}
 \end{figure}

\subsubsection{Quantitative comparison}
For this comparison, we consider data retrieved from the REST API of Yelp\footnote{\url{https://www.yelp.ie/dublin}}, as SERVICE-to-API does not apply to WebTables. We chose 
the Yelp API, as it is the only data source for which both systems offer the same functionality (our Twitter operator involves a microservice for performing sentiment analysis).
However, the findings of this experiment are representative of the general behaviour of the two systems with respect to any Web API.

For SERVICE-to-API, we stored data about businesses (burger joints in Chicago) in an RDF repository, because it
does not support queries that include API calls without triple patterns included in the query.
Then, we used the SERVICE keyword to join them with their names and IDs that are retrieved from the REST API of Yelp. Note that we used the original implementation of SERVICE-to-API, which was kindly provided to us by the authors of \cite{eswc18}. For \textsf{Ontop4theWeb}, we implemented a virtual table operator of Yelp and pose the SPARQL query Q1, which appears in Listing \ref{lst:yelp}, to retrieve the same data. 

SPARQL Query Q2 contains one more triple pattern (i.e., we also retrieve the rating of businesses) and it is described in Listing \ref{lst:yelp2}. There are different ways to express this query in the SERVICE-to-API, depending on the configuration of the repository. The closest definition 
seems to be the query shown in Listing \ref{lst:yelp5}, which is query  Q5. However, the fact that it returns different results suggests that this is not the case. Instead of returning the name and rating of the requested businesses, it returned the Cartesian product of all different burger businesses and all different rating values. So, if the SPARQL query Q2 is expected to return $|N|$ results, the query in Listing \ref{lst:yelp5} returns $|N \times S|$ results, where $|S|$ is the number of different rating values. We could briefly describe this phenomenon as \textit{a difference in semantics between SPARQL and the new language proposed~in~\cite{eswc18}}. 

Despite this significant difference between the two systems, 
we want to perform an exhaustive evaluation that highlights
their pros and cons.
To this end, we created and evaluated all different variations of 
configurations for the standard SPARQL queries Q1 and Q2.
We explain the differences in the SERVICE-to-API queries below. 

In SERVICE-to-API query Q1 (Listing \ref{lst:yelpq1}), we have stored  the names of businesses. 
So, we only need to retrieve the id's from the Yelp API. In SERVICE-to-API
query Q2 (Listing \ref{lst:yelpq2}), we want to retrieve both information from the Yelp API. In both cases,
we want to retrieve names and burger businesses in Chicago, so both queries are supposed to be equivalent to query Q1. However, these queries do not return the same results. SERVICE-to-API Query Q1 returns the same results as the standard SPARQL query Q1 that
was evaluated in our system, but SERVICE-to-API query Q2 returned many
false positives. These false positives were produced 
because the values that are bound to the variables involved in the query do not get joined, as in the case when the names are materialised in SERVICE-to-API query Q1. 

We did the same for SPARQL query Q2, which contains one more triple pattern
in its standard SPARQL representation.  Once we have at least one triple for each entity stored, we retrieve only the missing values using the SERVICE-to-API queries Q3 and Q4, which are described in Listings \ref{lst:yelp3} and \ref{lst:yelp4}, respectively. In this way, SERVICE-to-API returns the correct results, since the underlying triple store is forced to perform a JOIN between the materialized and the values that are returned from the API, instead of a Cartesian product. The trade-off, on the other hand, is that \textit{SERVICE-to-API cannot pose a query to retrieve the results directly through the API, as some form of materialization needs to be performed in order to retrieve correct results}.

SERVICE-to-API Query Q6 (Listing \ref{lst:yelp6}) differs from query Q5 only in that it uses the \texttt{BIND} operator instead of triple pattern (i.e., instead of storing the respective triple in a triple store). 
The reason why we performed this experiment was to execute a materialised-nothing query as the one that is performed in Ontop4theWeb query, where nothing is materialised in a database. The results of this query were eventually the same
as the results of the SERVICE-to-API query Q5. 

\begin{minipage}{0.47\textwidth}
\begin{lstlisting}[basicstyle=\ttfamily\scriptsize,frame=tb,caption={SPARQL query Q1},label={lst:yelp}]
select distinct ?id ?name  
where { 
 ?s yelp:name ?name .  ?s yelp:hasID ?id }
\end{lstlisting}
\end{minipage}
\begin{minipage}{0.47\textwidth}
\begin{lstlisting}[basicstyle=\ttfamily\scriptsize,frame=tb,caption={SPARQL query Q2},label={lst:yelp2}]
select distinct ?id ?name  
where { ?s yelp:name ?name . 
 ?s yelp:rating ?rating . 
 ?s yelp:hasID ?id }
\end{lstlisting}
\end{minipage}

\begin{lstlisting}[basicstyle=\ttfamily\scriptsize,frame=tb,caption={SERVICE-to-API query Q1 (eq. to SPARQL Q1) },label={lst:yelpq1}]
SELECT distinct  ?i ?name    
WHERE {
?x <http://www.w3.org/2000/01/rdf-schema#label> ?l . 
?x <http://yelp.com/ontology#name> ?name .
SERVICE <https://api.yelp.com/v3/businesses/{l}>{
( $.[\"id\"]) AS (?i)}} 
\end{lstlisting}

\begin{lstlisting}[basicstyle=\ttfamily\scriptsize,frame=tb,caption={SERVICE-to-API query Q2 (eq. to SPARQL Q1) },label={lst:yelpq2}]
SELECT distinct ?id ?name  WHERE {
?x <http://www.w3.org/2000/01/rdf-schema#label> ?l  SERVICE
<https://api.yelp.com/v3/businesses/
search?term=Burgers&location={l}&sort=2> 
($.[\"businesses\"][0:20][\"id\"],
$.[\"businesses\"][0:20][\"name\"]) AS (?id, ?name)}" 
\end{lstlisting}

\begin{lstlisting}[basicstyle=\ttfamily\scriptsize,frame=tb,caption={SERVICE-to-API query Q3 (eq. to SPARQL Q2) },label={lst:yelp3}]
SELECT distinct  ?i ?name ?rating    
WHERE {
?x <http://www.w3.org/2000/01/rdf-schema#label> ?l . 
?x <http://yelp.com/ontology#name> ?name .
SERVICE <https://api.yelp.com/v3/businesses/{l}>{
( $.[\"id\"],
$.[\"businesses\"][0:20][\"rating\"] ) AS (?id, ?r) }} 
\end{lstlisting}

\begin{lstlisting}[basicstyle=\ttfamily\scriptsize,frame=tb,caption={SERVICE-to-API query Q4 (eq. to SPARQL Q2) },label={lst:yelp4}]
SELECT distinct  ?i ?name    WHERE {
?x <http://www.w3.org/2000/01/rdf-schema#label> ?l . 
 ?x <http://yelp.com/ontology#name> ?name .
?x <http://yelp.com/ontology#rating> ?rating . 
SERVICE <https://api.yelp.com/v3/businesses/{l}>{
( $.[\"id\"]) AS (?i)}} 
\end{lstlisting}

\begin{lstlisting}[basicstyle=\ttfamily\scriptsize,frame=tb,caption={SERVICE-to-API query Q5 (eq. to SPARQL Q2)},label={lst:yelp5}]
SELECT distinct ?id ?b  WHERE {
?x <http://www.w3.org/2000/01/rdf-schema#label> ?l  
SERVICE
<https://api.yelp.com/v3/businessessearch?
term=Burgers&location={l}&sort=2> 
($.[\"businesses\"][0:20][\"id\"], 
$.[\"businesses\"][0:20][\"name\"],
$.[\"businesses\"][0:20][\"rating\"] 
AS (?id, ?b, ?r)}" 
\end{lstlisting}

\begin{lstlisting}[basicstyle=\ttfamily\scriptsize,frame=tb,caption={SERVICE-to-API query Q6 (eq. to SPARQL Q2) },label={lst:yelp6}]
 SELECT distinct ?id ?b ?r 
 WHERE {\n
 bind(\"Chicago\" as ?l) 
 SERVICE 
 <https://api.yelp.com/v3/businesses/
 search?term=Burgers&location={l}&sort=2
 { ($.[\"businesses\"][0:20][\"name\"],
 $.[\"businesses\"][0:20][\"name\"],
 $.[\"businesses\"][0:20][\"rating\"] )
 AS (?id, ?b, ?r)}" }
\end{lstlisting}

\textbf{Response time.} We evaluated these queries in both systems and we present the results in Figures \ref{fig:expyelp}  and \ref{fig:expyelpcalls}. 
The former depicts the query execution times and the latter the number of API calls invoked. In both cases, we consider warm and cold caches (on the left and right, respectively).
We observe that \textsf{Ontop4theWeb} is three times faster than SERVICE-to-API. The main reason is that \textsf{Ontop4theWeb} retrieves a set of tuples for each API call, which are then mapped into 
virtual RDF graphs. In contrast, SERVICE-to-API retrieves
one entry for each API call, yielding many more API calls in order to get the same information. 

Another observation is that \textsf{Ontop4theWeb} by design benefits more from caching than the system in comparison. 
We cache 
the entire table for each API call, while SERVICE-to-API performs an API call for each tuple, which means that only one tuple is cached each time.
Hence, for a result set consisting of $N$ tuples, 
\textsf{Ontop4theWeb} will cache the entire result set as a virtual table that is retrieved from a single API call. In contrast, SERVICE-to-API 
needs at least $N$ calls, of which at most one will be cached.

One could argue that the comparison between the two systems might not  be fair, as it seems that the two systems are have differences (e.g., they implement different languages). 
However, our motivation for these experiments were to compare the performance and functionality with a system that offers similar functionality, answering to the following question "If \textsf{Ontop4theWeb} was not in place, what would be the system that we would use in order to have similar functionality".
SERVICE-to-API was the only alternative in this direction. 

Another argument could be the fact that, given implementation internals of   SERVICE-to-API (e.g., retrieving tuples instead of virtual tables), 
the results of the experimental evaluation are reasonable. However, 
these internal implementation details were not obvious until we executed the experiments. Our experiments highlighted these issues and led us to discover these differences in the design and implementation that are the cause of the performance results presented above.

\textbf{Accuracy.}
We now investigate how accurate are the results
in both systems when posing the queries described in the previous section. 
Table \ref{table:precision} shows how precise were the results returned
by the two systems in comparison, Table \ref{table:recall} presents
the recall, Table \ref{table:accuracy} presents the accuracy metrics 
and Table \ref{table:f1} presents the F1-score of all six queries
that were evaluated. 

We observe that we make is that both systems
do not produce false negatives, so the recall is always 1. 
SERVICE-to-API produces false positives that reduce the system's precision and accuracy and, inevitably, its F1-score. The reason is that, 
as discussed in previous sections, it 
returns the cartesian product of the bindings of all variables involved in
an API call. 

\textbf{Summary.}
The findings of this experiment show that not only is \textsf{Ontop4theWeb} more efficient
in terms of response time in comparison with the current state-of-the-art, but it also
produces accurate results in all cases. 
Our experiments 
demonstrate that in order to obtain correct results from SERVICE-to-API,
one needs to
partially store the data and use a REST API complementarily. Even in this case, however, the functionality that is offered is a subset of the functionality that is offered by our system, while the execution time of queries, even with all optimisations enabled, is considerably larger. 

\label{subsec:accuracy}
\begin{table}[h]
\begin{tabular}{|c|c|c|c |c|c|c|} 
 \hline
  System & Q1 & Q2 & Q3 & Q4 & Q5 & Q6 \\
  \hline
 SERVICE-to-API & 1 & 0.05 & 0.05 & 1 & 0.016 & 0.016  \\ 
 Ontop4TheWeb   & 1 & 1 & 1 & 1 & 1 & 1 \\
 \hline
\end{tabular}
 \caption{Precision}
\label{table:precision}
\end{table}

\begin{table}[h]
\begin{tabular}{|c|c|c|c |c|c|c|} 
 \hline
  System & Q1 & Q2 & Q3 & Q4 & Q5 & Q6 \\
  \hline
 SERVICE-to-API & 1 & 1 & 1 & 1 & 1 & 1  \\ 
 Ontop4TheWeb   & 1 & 1 & 1 & 1 & 1 & 1 \\
 \hline
\end{tabular}
 \caption{Recall}
\label{table:recall}
\end{table}

\begin{table}[h]
\begin{tabular}{|c|c|c|c |c|c|c|} 
 \hline
  System & Q1 & Q2 & Q3 & Q4 & Q5 & Q6 \\
  \hline
 SERVICE-to-API & 1 & 0.05 & 0.05 & 1 & 0.016 & 0.016  \\ 
 Ontop4TheWeb   & 1 & 1 & 1 & 1 & 1 & 1 \\
 \hline
\end{tabular}
 \caption{Accuracy}
\label{table:accuracy}
\end{table}

\begin{table}[h]
\begin{tabular}{|c|c|c|c |c|c|c|} 
 \hline
  System & Q1 & Q2 & Q3 & Q4 & Q5 & Q6 \\
  \hline
 SERVICE-to-API & 1 & 0.09 & 0.09 & 1 & 0.015 & 0.015  \\ 
 Ontop4TheWeb   & 1 & 1 & 1 & 1 & 1 & 1 \\
 \hline
\end{tabular}
 \caption{F1-score}
\label{table:f1}
\end{table}

\section{Conclusions }
\label{sec:conclusions}

This paper presents \textsf{Ontop4theWeb}, a novel system for querying Web data on-the-fly using SPARQL. \textsf{Ontop4theWeb} extends SQL with virtual table operators, embeds them into mappings and makes an OBDA system compliant to them. Our extensive experimental evaluation verified that \textsf{Ontop4theWeb} goes beyond the state of the art, not only in terms of functionality, but also in terms of performance. Our approach complements traditional approaches of querying data using SPARQL, accommodating  
the Variety and Velocity of Web data.  

In the future, we will use our system as a framework to solve more research problems that include data analysis tasks and make the results available as virtual RDF triples on-the-fly. We will also exploit the extensibility of our system to support more use cases, such as creating virtual RDF graphs on top of XML documents.

\bibliographystyle{ACM-Reference-Format}
\bibliography{bibliography.bib}

\end{document}